\documentclass[12pt]{iopart}
\usepackage{iopams}
\usepackage[utf8]{inputenc}
\usepackage{graphicx}
\usepackage{siunitx}
\usepackage{xspace}
\usepackage{hyperref}
\usepackage{amsfonts}

\expandafter\let\csname equation*\endcsname\relax

\expandafter\let\csname endequation*\endcsname\relax

\usepackage{amsmath} 
\usepackage{amssymb}
\usepackage{braket}
\usepackage{dsfont}
\usepackage{xcolor}

\newcommand{\BePlus}{$^9$Be$^+$\xspace}

\usepackage{letltxmacro}
\LetLtxMacro{\ORIGselectlanguage}{\selectlanguage}
\makeatletter
\DeclareRobustCommand{\selectlanguage}[1]{%
	\@ifundefined{alias@\string#1}
	{\ORIGselectlanguage{#1}}
	{\begingroup\edef\x{\endgroup
			\noexpand\ORIGselectlanguage{\@nameuse{alias@#1}}}\x}%
}
\newcommand{\definelanguagealias}[2]{%
	\@namedef{alias@#1}{#2}%
}
\makeatother
\definelanguagealias{en}{english}
\definelanguagealias{En}{english}
\definelanguagealias{de}{german}
\definelanguagealias{EN}{english}


\begin{document}
	\title[Numerical optimization of amplitude-modulated pulses]{Numerical optimization of amplitude-modulated pulses in microwave-driven entanglement generation}
	
	\author{M.~Duwe$^{\star,1,2}$, G.~Zarantonello$^{\star,\dagger,^{1,2}}$, N.~Pulido-Mateo$^{1,2}$, H.~Mendpara$^{1,2}$, L.~Krinner$^{1,2}$, A.~Bautista-Salvador$^{1,2,3}$, N.~V.~Vitanov$^{4}$, K.~Hammerer$^{5}$, R.~F.~Werner$^{6}$ and C.~Ospelkaus$^{\ddag,1,2,3}$ }
	\address{$^1$ Institut für Quantenoptik, Leibniz Universität Hannover, Welfengarten 1, 30167 Hannover, Germany}
	\address{$^2$ Physikalisch-Technische Bundesanstalt, Bundesallee 100, 38116 Braunschweig, Germany}
	\address{$^3$ Laboratorium für Nano- und Quantenengineering, Leibniz Universität Hannover, Schneiderberg 39, 30167 Hannover, Germany}
	\address{$^4$ Department of Physics, St. Kliment Ohridski University of Sofia, 5 James Bourchier blvd, 1164 Sofia, Bulgaria}
	\address{$^5$ Institut für Theoretische Physik und Institut für Gravitationsphysik (Albert-Einstein-Institut), Leibniz Universität Hannover, Appelstrasse 2, 30167 Hannover, Germany}
	\address{$^6$ Institut für Theoretische Physik, Leibniz Universität Hannover, Appelstrasse 2, 30167 Hannover, Germany}

	\begin{abstract}
		Microwave control of trapped ions can provide an implementation of high-fidelity two-qubit gates free from errors induced by photon scattering. Furthermore, microwave conductors may be embedded into a scalable trap structure, providing the chip-level integration of control that is desirable for scaling. Recent developments have demonstrated how amplitude modulation of the gate drive can permit a two-qubit entangling operation to become robust against motional mode noise and other experimental imperfections. Here, we discuss a method for the numerical optimization of the microwave pulse envelope to produce gate pulses with improved resilience, faster operation and higher energy efficiency.
	\end{abstract}
	

\noindent$^\star$MD and GZ contributed equally to this work.\\
\noindent$^\dagger$Present address: National Institute of Standards and Technology, Boulder, Colorado 80305, USA\\
\noindent$^\ddag$christian.ospelkaus@iqo.uni-hannover.de\\

	\maketitle
	
	\section{Introduction}
 Trapped ions are a leading scalable platform for the implementation of quantum algorithms~\cite{bermudez_assessing_2017,bruzewicz_trapped-ion_2019,pino_demonstration_2021}. Scaling any hardware will ultimately require the implementation of quantum error correction codes~\cite{steane_error_1996,knill_quantum_2005} to prevent error propagation in large-scale algorithms. This will require quantum gates with fidelities beyond the fault-tolerance threshold~\cite{preskill_john_reliable_1998,knill_physics:_2010}. A universal set of quantum gates requires single-qubit gates and one two-qubit gate capable of entanglement generation~\cite{divincenzo_two-bit_1995}. For trapped ions, single-qubit operations have already reached error rates well below $10^{-4}$~\cite{brown_single-qubit_2011,harty_high-fidelity_2014}. A major experimental challenge is to obtain similar error rates for a two-qubit entangling gate. Experimental results are approaching the desired gate fidelity~\cite{ballance_high-fidelity_2016,gaebler_high-fidelity_2016,srinivas_high-fidelity_2021,clark_high-fidelity_2021} where large-scale error correction could reasonably be implemented~\cite{ryan-anderson_realization_2021,hilder_fault-tolerant_2021}.
	Unfortunately, two-qubit entangling gates can be affected by a variety of imperfections, the seriousness of which depend on the type of gate itself. Quantum control methods allow to analytically or numerically improve the performance of the gate, providing resilience and robustness against specific error sources. The specifics of the protocol implemented depend on the source of errors addressed for the gate.
	In the case of noise connected to the qubit frequency, the most straightforward control protocol is the Hahn echo~\cite{hahn_spin_1950}, but depending on the gate protocol more advanced schemes can be required. There are multiple protocols which employ pulsed dynamic decoupling~\cite{manovitz_fast_2017,arrazola_pulsed_2018} and continuous dynamic decoupling~\cite{bermudez_robust_2012} or other forms of error suppression~\cite{ivanov_composite_2015,arrazola_hybrid_2020}.
	In the case of errors connected to the ion's state of motion, several methods have been studied: Walsh modulation~\cite{hayes_coherent_2012}, multi-tone fields~\cite{haddadfarshi_high_2016,shapira_robust_2018,webb_resilient_2018}, phase modulation~\cite{milne_phase-modulated_2018}, frequency modulation~\cite{leung_entangling_2018} and amplitude modulation~\cite{zhu_trapped_2006,roos_ion_2008}. The latter has been extensively used in laser-driven operations~\cite{zhu_trapped_2006,roos_ion_2008,schafer_fast_2018,figgatt_parallel_2019}, and more recently demonstrated for microwave-driven operations~\cite{zarantonello_robust_2019}. 
	Hybrid schemes to provide simultaneous insensitivity to motional mode and qubit frequency instabilities have been proposed~\cite{lishman_trapped-ion_2020,sutherland_laser-free_2020}.\\
	
	Here we present a method to perform numerical optimization of a pulse envelope for amplitude-modulation in M{\o}lmer-S{\o}rensen entangling gates. The method described here allows the area enclosed in phase-space trajectories of the ion motion to be insensitive to trap and pulse parameters, allowing faster operations while maintaining the previously demonstrated insensitivity. The energy used for the gate is minimized, an important feature in the case of microwave driven operations~\cite{mintert_ion-trap_2001,ospelkaus_trapped-ion_2008,sutherland_versatile_2019}, especially in cryogenic environments where cooling power should be limited~\cite{srinivas_laser-free_2020,dubielzig_ultra-low_2021}, as excess energy could affect the stability of the mode.
	
	\section{Numerical optimization}
	
	\begin{figure*}[tb]
		\centering
		\includegraphics[width=\textwidth]{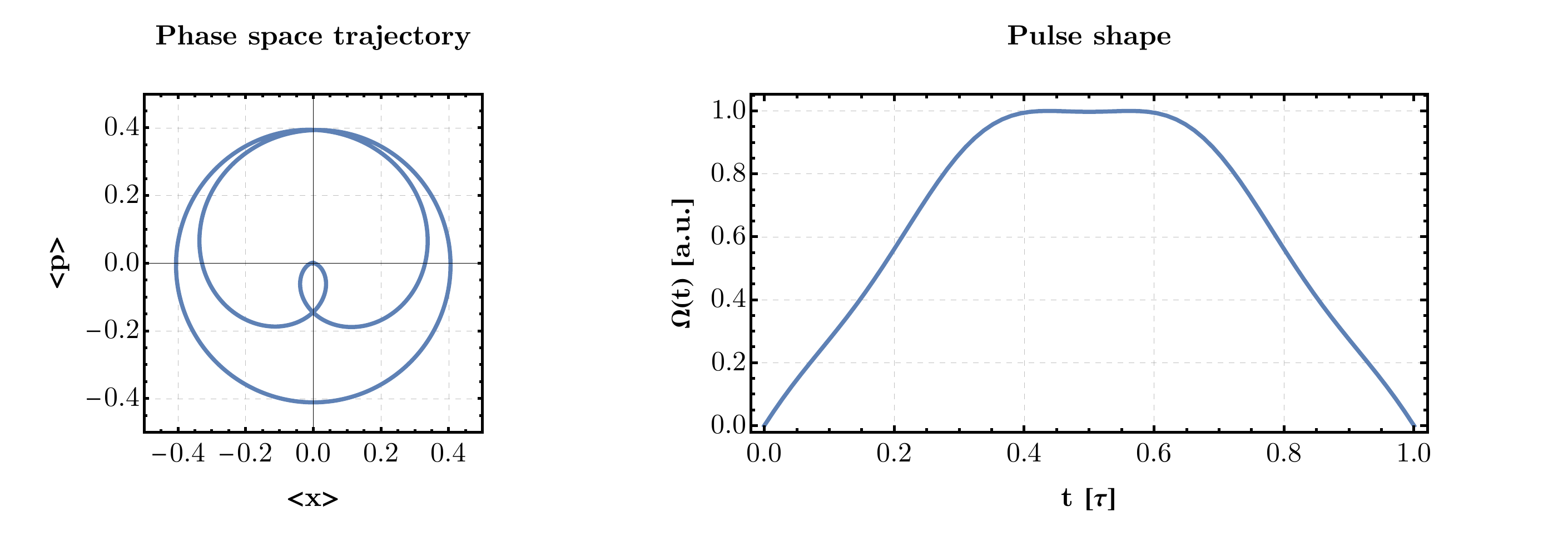}
		\caption{Pulse shape and phase space trajectory for $\delta \tau= 6\pi$. The gate time is $\tau=1000.4\SI{}{\micro\second}$ with a detuning of $\delta=2.998$ kHz. These values are obtained for a specific Rabi frequency $\Omega_{MS}/2\pi=1.18\SI{}{\kilo\hertz} $.}
		\label{fig:6pi}
	\end{figure*}
	
	\label{sec:num_opt}
	The numerical method presented here aims to optimize the amplitude of the bichromatic microwave field which drives the M{\o}lmer-S{\o}rensen  entangling gate~\cite{molmer_multiparticle_1999,solano_deterministic_1999,milburn_ion_2000}. The gate dynamics can be described by introducing a basic control function \(\Omega :[0,\tau]\xrightarrow{}\mathbb{R}\). Such functions modulate the motion in phase space of the harmonic oscillator describing the ions' secular motion
	
	\begin{equation}\label{eq:1}
		\begin{split}		
			p(t)= \int_0^t ds \ \sin(\delta s)\Omega(s) \\
			q(t)= \int_0^t ds \ \cos(\delta s)\Omega(s)~,
		\end{split}
	\end{equation}
	where $\delta$ is the detuning from the motional mode frequency of the bichromatic drive. At a specific time $\tau$, the gate time, the trajectory described in~\eqref{eq:1} should ideally constitute a loop with an enclosed area $|A|=\pi/2$. In case the trajectory does not return to the initial phase space position or the enclosed area differs from $\pi/2$ the gate fidelity $\mathcal{F}$ will be affected by an error. As a fidelity measure we use the overlap of the generated state with the target Bell state. Other measures exist, and should eventually be considered too. In terms of \(\Omega\) the area is
	
	\begin{equation}\label{eq:3}
		\begin{split}		
			A&=\int pdq =\int_0^\tau dt\ \cos(\delta t)\Omega(t) \int_0^t ds\ \sin(\delta s)\Omega(s)\\
			&=\frac{1}{2}\int_0^\tau dt \int_0^\tau ds \ \Omega (t)K(t,s)\Omega(s)
		\end{split}
	\end{equation}
	with the kernel 
	
	\begin{equation}\label{eq:4}
		K(t,s)=\cos(\delta \max(t,s)) \sin(\delta \min(t,s))~,  
	\end{equation}	
	with $t,s \in [0,\tau].$ Furthermore, as previously stated, it is of interest to minimize the energy dissipated in the trap. Since the gate Rabi rate is proportional to the current flowing in the trap conductors, the energy will be proportional to
	
	\begin{equation}\label{eq:5}
		E=\int_0^\tau dt\ \Omega(t)^2.
	\end{equation}
	The optimization done here generates pulses that have a fixed area in phase space while minimizing the energy. Since both quantities are given by quadratic expressions, this amounts to finding the pulse $\Omega$ which satisfies the generalized eigenvalue equation
 	\begin{equation}
 		\hat{A}\Omega=\lambda \hat{E}\Omega
 	\end{equation}
    with the largest possible eigenvalue $\lambda$. Here $\hat{A}$ and $\hat{E}$ are the operators representing the quadratic expressions $A$ and $E$. That is $\langle\Omega|\hat{A}|\Omega\rangle=A$, so $\hat{A}$ is given by the above kernel $K$, and $\hat{E}$ is the identity operator. This general form will also apply in the discretized version that we solve numerically, although $\hat{E}$ will then be more complicated. One can use the scheme to achieve additional desirable features, for example, smoothness. To that end we will add to $E$ a small penalty term with the norm square of $\Omega'$, i.e., use a so-called Sobolev norm. Moreover, it is easy to include arbitrary linear constraints $\langle\phi,\Omega\rangle=0$, by solving the generalized eigenvalue problem on a subspace. The most important of these is that the loop closes in phase space, i.e., $p(\tau)=q(\tau)=0$ in \ref{eq:1}. We can also ensure that small variations in the detuning $\delta$ to not disturb the loop closure, by demanding that the derivatives of $p(\tau)$ and $q(\tau)$ with respect to $\delta$ vanish as well. That is we demand $\Omega$ to be orthogonal to the following four functions    
    \begin{alignat*}{10}
    	\phi^1&=  \cos(\delta t),\ {} &\phi^2&=\sin(\delta t)\label{eq:excluded}\\
    	\phi^3&=  t \sin(\delta t),\ &\phi^4&= t \cos(\delta t)\nonumber~.\\
    \end{alignat*}
    $\phi^1$ and $\phi^2$ ensure that $p(\tau)=q(\tau)=0$ because 
        \begin{equation}\braket{\phi^1,\Omega}=\int_0^\tau dt\cos(\delta t) \Omega(t)=0\end{equation} and
    \begin{equation}\braket{\phi^2,\Omega}=\int_0^\tau dt\sin(\delta t) \Omega(t)=0.\end{equation} Therefore they ensure a closed loop in phase space. The functions \(\phi^3\) and \(\phi^4\) are the derivatives of $\phi^1$ and $\phi^2$ with respect to $\delta$ and ensure that small variations in the detuning leave the loop unchanged. Potentially the amount of excluded functions can be increased. One might consider to include even higher derivatives of $\phi^1$ and $\phi^2$ for more stability with respect to the detuning in higher derivatives. We restricted ourselves to these four functions to not make the subspace too large, which would come at the expense of increased energy. This derivative strategy is reminiscent of the approach used to construct composite pulses. However, here it is used to construct the shape of the pulse rather than the phases of the composite sequence of pulses, with the benefit of having much shorter driving field duration.
    
	For a numerical analysis, the gate time interval is discretized into \(n+1\) pieces, with cut points \(0~\SI{}{\second}=:t_0<t_1<t_2<...<t_n<t_{n+1}:=\tau.\) In the optimization, various functions have to be evaluated at these points. The following basis is used for $\Omega$
	\begin{equation}
		\chi_k(t)=  \begin{cases}
			\begin{array}{rcl}\frac{t-t_{k-1}}{t_k-t_{k-1}} &,& t_{k-1}<t\leq t_k 
				\\\frac{t_{k+1}-t}{t_{k+1}-t_{k}} &,& t_k<t\leq t_{k+1}
				\\0 &,& \text{otherwise.}  
		\end{array}	\end{cases} 
	\end{equation}	
	Note that $ \chi_k(t_j)=\delta_{kj}$, so the expansion coefficients in
	
	\begin{equation}\label{eq:ref}
		\Omega(t)=\sum_{k=1}^n\omega_k \chi_k(t)
	\end{equation}	
	are exactly the values \(\omega_k=\Omega(t_k)\). These coefficients give the gate Rabi rate value at a specific moment $t_k$. When \(\phi^{(i)}\) are the excluded functions, the projection onto their complement is
	
	\begin{equation}\label{eq:p}
		p=\mathds{1} -\sum_{ij}\ket{\phi^{(i)}}G_{ij}^{-1}\bra{\phi^{(j)}}~,
	\end{equation}
	where $G_{ij}=\braket{\phi^{(i)},\phi^{(j)}}$ is the corresponding Gram matrix, and \(G^{-1}\) is it's matrix inverse. The energy kernel $\hat{E}_{ij}=\braket{\chi_i,\chi_j}$ of~\eqref{eq:5} is adjusted with the Sobolev norm so that
		\begin{equation}
		\hat{E}_{ij}=\braket{\chi_i,\chi_j}+c\cdot \braket{\chi_i',\chi_j'}
	\end{equation}	
	to ensure a soft start, necessary to avoid pseudopotential kicks from parasitic electric fields~\cite{warring_techniques_2013}. In the following we restrict ourselves to the case $c=1$. This specific choice was made in order to ensure a soft start without increasing the energy too much. If $c=0$ then a soft start of the pulse is not guaranteed.

	The objective is to minimize the energy $E$ while keeping the area $A$ constant. This can be achieved by solving the generalized eigenvalue problem $\hat{\mathcal{A}}\Omega=\lambda \hat{\mathcal{E}}\Omega$ where $\hat{\mathcal{A}}=p\hat{A}p$ and $\hat{\mathcal{E}}=p\hat{E}p$ are the kernels mentioned above but projected onto the subspace with the projector $p$ defined earlier~\eqref{eq:p}. The eigenvector with the largest eigenvalue provides the coefficients $\omega_k$ that have the best ratio of energy to area. Higher orders are therefore not considered. The ratio of $\delta$ and $\tau$ defines the corresponding eigenvector because the area kernel is constructed for a given detuning. Therefore for all detunings the area kernel is different. This leads to diverse eigenvectors. For the M{\o}lmer-S{\o}rensen gate it needs to fulfill $\delta \cdot \tau=K \cdot 2\pi$ where $K$ is an integer number. The possibility to perform a gate starts with $K=1$.  A possible eigenvector and its phase space trajectory is shown in Fig.~\ref{fig:6pi}. \\
	
\begin{figure}[tb]
		\centering
		\includegraphics[width=\columnwidth]{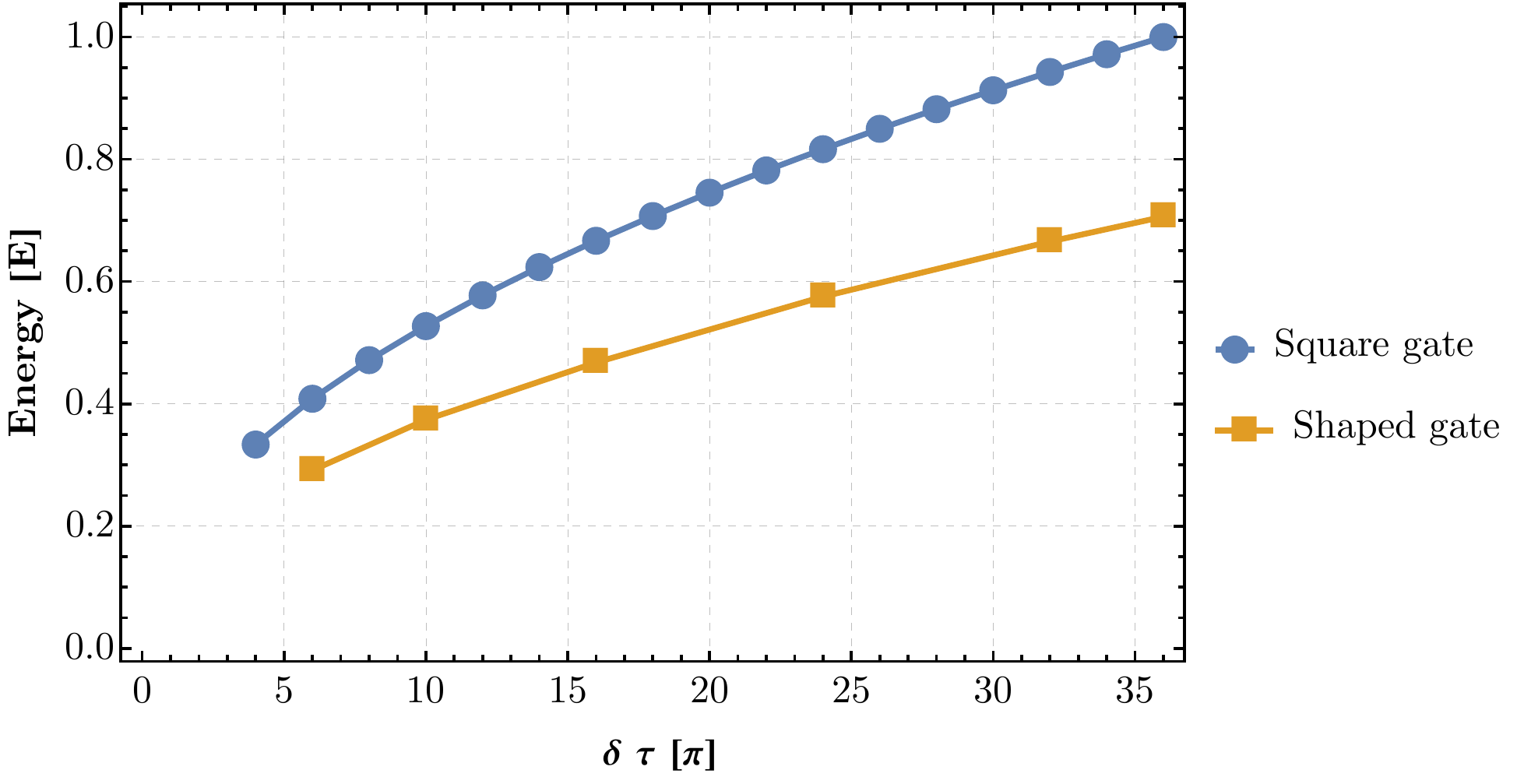}
		\caption{Calculated energy dissipated in the trap for different values of $\delta \tau$. Blue dots correspond to the square gate and orange to the optimized shapes. The energy dissipated depends on the actual experimental conditions, the values reported here are normalized for comparison.}
		\label{fig:energycomparison}
	\end{figure}

	Fig.~\ref{fig:energycomparison} shows the energy dissipated in the trap for the square and the shaped gates. The energy $E$ is calculated with~\eqref{eq:5} for both shapes. For the same value of $\delta \tau$, the optimized pulses dissipate approximately 25$\%$ less than the square pulses energy. The main reason for this is the soft start of the pulses, which lowers the integral of $\Omega(t)^2$ over the gate time $\tau$. The largest energy corresponds to the square pulse with $\delta \tau =36 \pi$. All energies in Fig.~\ref{fig:energycomparison} are normalized to this value. The dissipated energy contributes to the gate errors because it leads to motional instability. Therefore pulses that dissipate less energy are preferred to minimize effects such as the mode frequency chirp~\cite{hahn_integrated_2019}. \\
	
	\section{Experimental demonstration}
	\label{sec:exp}
	The experimental demonstration of the entangling gate pulse envelopes obtained using the method described in Sec.~\ref{sec:num_opt} has been done in the setup described in~\cite{hahn_integrated_2019,hahn_two-qubit_2019}. The experiments have been performed on \BePlus ions at a static magnetic field of $\left|\mathrm{\bf{B_0}}\right|=22.3\,$mT, where the chosen qubit transition, $\mathrm{^{2}S_{1/2}}\left|2,1\right>$ $\leftrightarrow$ $\mathrm{^{2}S_{1/2}}\left|1,1\right>$, is first-order field independent. The enhanced frequency stability of the transition is reflected in a long coherence time of the qubit~\cite{langer_long-lived_2005}. Doppler cooling and detection is performed with a $313\,$nm laser resonant with the $\mathrm{^{2}S_{1/2}}\left|F=2,m_F=2\right>$ $\leftrightarrow$ $\mathrm{^{2}P_{3/2}}\left|m_J=\frac{3}{2},m_I=\frac{3}{2}\right>$ transition. For integrated microwave control, three conductors are embedded in the surface-electrode trap: two conductors for driving carrier transitions and one for sideband operations. The latter is designed~\cite{carsjens_surface-electrode_2014} to produce a strong magnetic field gradient optimized for spin-motional coupling. The microwave amplitude modulation setup is described in~\cite{zarantonello_robust_2019}. For implementation purposes optimized pulse envelopes are produced with $\delta \tau =$ $6\pi$, $10\pi$, $18\pi$, $24\pi$ and $36\pi$. For a maximum gate Rabi rate of $\Omega_\mathrm{MS}/2\pi=1.18\,$kHz, the pulse durations are $1000.4~\SI{}{\micro\second}$, $1324.8~\SI{}{\micro\second}$, $1793.77~\SI{}{\micro\second}$, $2083.4~\SI{}{\micro\second}$ and $2548.7~\SI{}{\micro\second}$.\\
	
\begin{figure}[tb]
		\centering
		\includegraphics[width=\columnwidth]{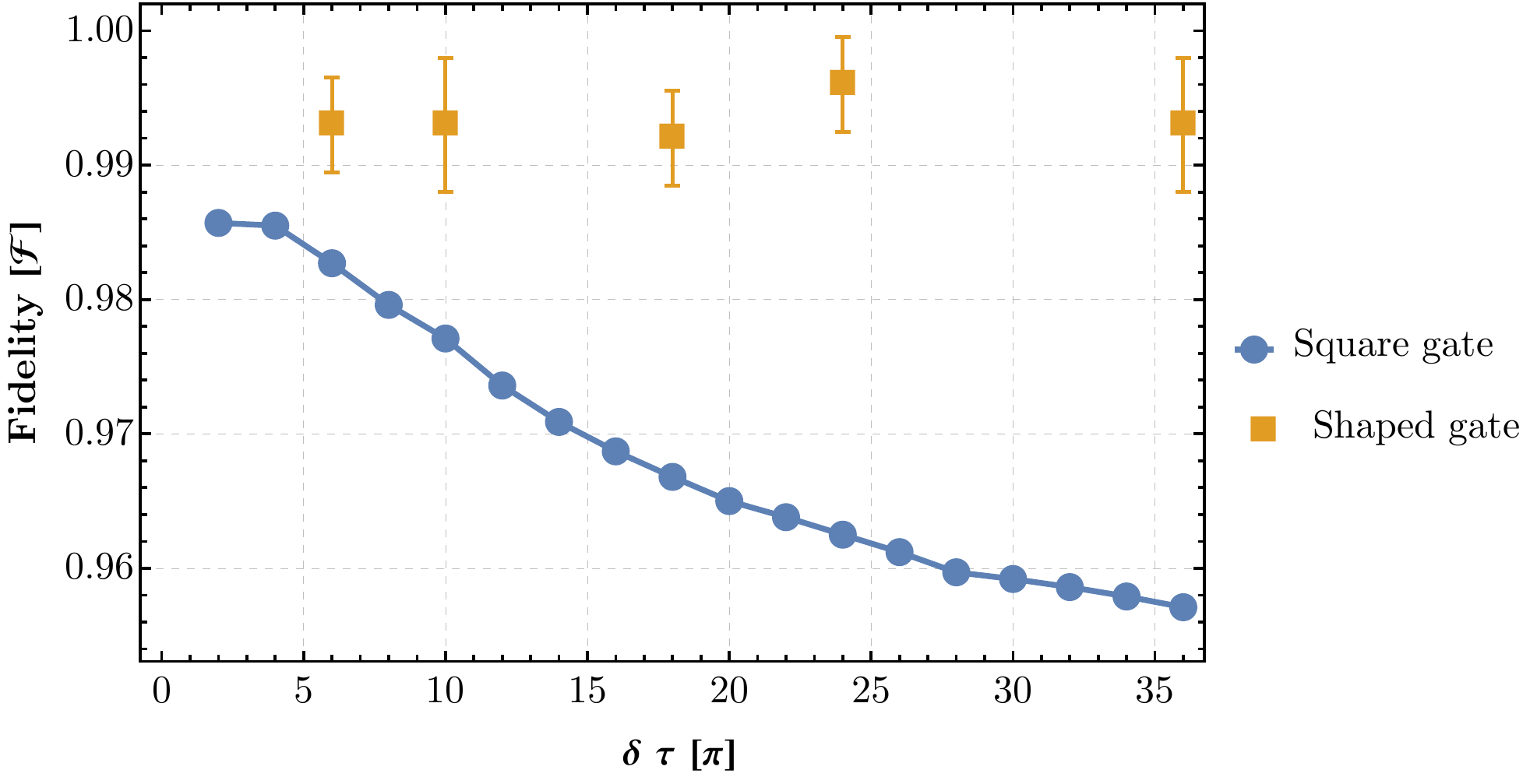}
		\caption{Entangled state fidelity for different values of $\delta\tau$. Blue circles represent an estimation using an error model for square-pulse gates. Orange squares represent measured values for different amplitude-modulated gates generated by numerical optimization. }
		\label{fig:comparison}
	\end{figure}

	The maximally entangled states generated have been analyzed with a partial tomography procedure~\cite{sackett_experimental_2000}. All state populations have been estimated from the global fluorescence emission of the ions by using appropriately placed thresholds in the photon count histograms. Experimental results are reported in Fig.~\ref{fig:comparison} and compared with the expected theoretical performance of a standard square-pulse gate. $\pi$ and $\pi / 2$ rotations used for state preparation, shelving and analysis are implemented using composite pulses. Specifically, the U5a sequence has been used for $\pi$ pulses~\cite{genov_correction_2014} and the $5$ pulse sequence for the $\pi/2$ analysis pulse~\cite{torosov_smooth_2011}.
	
	The fidelities for simulated square-pulse M{\o}lmer-S{\o}rensen gates have been obtained using the error model developed in~\cite{hahn_integrated_2019} and adjusted to reflect the experimental conditions. For the addressed out-of-phase radial motional mode, we considered an average motional state occupation of $\bar{n}=0.4$, a heating rate of $ \dot{\bar{n}} = 8\,\mathrm{s}^{-1}$ and an intrinsic linewidth of $ 2\pi \times 61\,\mathrm{Hz} $. A single spectator in-phase mode was included, detuned by $\delta_s/2\pi=96\,$kHz with $\bar{n}=1$. The experiments have been performed without a warm-up pulse employed in earlier experiments to minimize motional mode frequency fluctuations, called "frequency chirp", induced by thermal transients in the trap~\cite{hahn_integrated_2019}. The absence of this thermalization process before the gate means that the full frequency chirp has to be included in the simulation with a ramp of $0.3\,\SI{}{\hertz /\micro\second}$ for up to $1000\,$\SI{}{\micro\second}. For each simulated point, the gate detuning $\delta$ has been optimized to yield the highest resulting fidelity. Parameters regarding the imperfection of the microwave pulse shape have been left as in the original model. We place a lower bound of 0.5 s on the internal-state coherence time. Variations of the qubit transition frequency have not been included as an error source in the simulation. The reason behind this choice is that ideally the gate is performed at the same point of minimal AC-Zeeman shift (ACZS) as for the amplitude-modulated gates. Note that despite this choice, it is expected that fast variations, or non-perfect calibrations, of the ACZS could lead to errors in AM gates. Further details of the error model are described in~\cite{schulte_entanglement_2020}.
	The results shown in Fig.~\ref{fig:comparison} consistently show infidelities below $1\%$. The large difference in the fidelity between amplitude-modulated and square-pulse gates at the highest values of $\delta\tau$ demonstrates the resilience of the optimized gates.\\
		
	\section{Conclusions}
	\label{sec:res}
	The numerical method described here provides the capability to generate optimized amplitude-modulated gates that are robust with respect to disturbances of the motional mode. In addition, the algorithm minimizes the energy per pulse used to generate a maximally entangled Bell state in microwave driven operations. The experimental verification has been done by implementing multiple optimized pulse envelopes, consistently demonstrating an infidelity in the $10^{-3}$ range. All gates produced with this method were faster than the one described in our previous work~\cite{zarantonello_robust_2019}, the fastest of which had $\tau\approx1000$~\SI{}{\micro\second}.
	
	In the future, such shaped gates can be integrated with other protocols, such as dynamic decoupling, to suppress gate errors connected to the ACZS affecting the qubit frequency during gate operation. The integration of these decoupling protocols would ease the requirements concerning the minimization of the ACZS and provide resilience to small changes of the ACZS.
	To further increase the gate speed, it is necessary to increase the magnetic field gradient driving the gate. One possibility is given by advanced three-dimensional microwave structures~\cite{hahn_multilayer_2019}.
	Given the large errors resulting from the global detection of two-ion fluorescence, more advanced schemes of error characterization are required, possibly in a computational contest, such as benchmarking methods~\cite{gaebler_randomized_2012,erhard_characterizing_2019,baldwin_subspace_2020}.

	\ack
		We thank M.~Schulte for participating in the early stages of the project. We thank P.~O.~Schmidt and S.~A.~King for helpful discussions. We acknowledge funding from the European Union Quantum technology flagship under project `MicroQC', from `QVLS-Q1' through the VW foundation and the ministry for science and culture of Lower-Saxony, from the Deutsche Forschungsgemeinschaft (DFG, German Research Foundation) under Germany’s Excellence Strategy - EXC-2123 QuantumFrontiers - 390837967 and through  the collaborative research center SFB 1227 DQ-\textit{mat}, projects A01, A05 and A06, and from PTB and LUH.\\

\section*{References}

\bibliography{qc}
\bibliographystyle{iopart-num}

\end{document}